\documentclass[11pt]{iopart}
\pdfoutput=1
\expandafter\let\csname equation*\endcsname\relax
\expandafter\let\csname endequation*\endcsname\relax
\usepackage{amsmath}
\usepackage{amssymb}
\usepackage{color}
\usepackage{graphicx}
\usepackage[sort&compress,numbers]{natbib}
\usepackage{verbatim}
\usepackage{amsfonts}
\usepackage{braket}
\usepackage{dsfont}
\usepackage[compat=1.1.0]{tikz-feynman}
\usepackage[english]{babel}
\usepackage{hyperref}
\hypersetup{urlcolor=red, colorlinks=true}  
\usepackage{pgfplots}
\usepackage{url}
\pgfplotsset{compat=1.12}
\usepackage{setspace}
\usepackage{subcaption}
\usepackage{slashed}

\begin{document}

\title[On critical properties of Berry curvature in Kitaev Honeycomb model]{On critical properties of Berry curvature in Kitaev Honeycomb model}

\author{Francesco Bascone$^{1,2}$, Luca Leonforte$^{3}$,  Davide Valenti$^{3,4}$, Bernardo Spagnolo$^{3,5,6}$ and Angelo Carollo$^{3,5}$}

\address{$^1$Dipartimento di Fisica "E. Pancini", Universit\`{a} di Napoli Federico II, Complesso Universitario di Monte S. Angelo Edificio 6, via
Cintia, 80126 Napoli, Italy}
\address{$^2$Istituto Nazionale di Fisica Nucleare, Sezione di Napoli, Complesso Universitario di Monte S. Angelo Edificio 6, via
Cintia, 80126 Napoli, Italy}
\address{$^3$Dipartimento di Fisica e Chimica ``Emilio Segr\'{e}", Group of Interdisciplinary Theoretical Physics, Universit\`{a} di Palermo, Viale delle Scienze, 
Edificio 18, I-90128 Palermo, Italy}
\address{$^4$IBIM-CNR Istituto di Biomedicina ed Immunologia Molecolare
``Alberto Monroy", Via Ugo La Malfa 153, I-90146 Palermo, Italy}
\address{$^5$Radiophysics Department, National Research Lobachevsky State University of Nizhni Novgorod, 23 Gagarin Avenue, Nizhni Novgorod 603950, Russia}
\address{$^6$Istituto Nazionale di Fisica Nucleare, Sezione di Catania, Via
S. Sofia 64, I-90123 Catania, Italy }
\ead{francesco.bascone@na.infn.it
}

\begin{abstract}
{\color{black} We analyse the Kitaev honeycomb model, by means of
the Berry curvature with respect to Hamiltonian parameters. We
concentrate on the ground-state vortex-free sector, which allows us
to exploit an appropriate Fermionisation technique. The parameter
space includes a time-reversal breaking term which provides an
analytical headway to study the curvature in phases in which it
would otherwise vanish. The curvature is then analysed in the limit
in which the time-reversal-symmetry-breaking perturbation vanishes.
This provides remarkable information about the topological phase
transitions of the model}. A non-critical behaviour is found in the
Berry curvature itself, which shows a {\color{black} distinctive}
behaviour in the different phases. The analysis of the first
derivative shows a critical behaviour around the transition point.
\end{abstract}


\vspace{2pc}

\submitto{\JSTAT}

\maketitle

\section{Introduction}
\label{intro}

Topological phase transitions (TPTs) have emerged as a new paradigm,
since they do not fall under the Landau theory description, where
phases are characterised by local order parameters and symmetry
breaking occurring across criticalities. Topological phases indeed
are identified in the bulk by topological invariants, i.e.
quantities that only depend on the topology, that are constructed
out of ground states
properties~\cite{Altland1997,Schnyder2008,Ryu2010,Chiu2016}.
Topological systems have attracted intense interest because of their
peculiar properties, such as topologically protected edge
states~\cite{Hatsugai1993} or exotic statistics
excitations~\cite{Laughlin1983,Arovas1984,Nayak2008}. Moreover, new
materials are discovered and experiments are performed in the
direction of probing anyonic excitations, which have promising
applications in many other areas such as quantum computing.

Most of the literature on TPTs concerns the zero temperature case,
where the systems are described by pure states. Recently efforts
were also made in the direction of a mixed state
{\color{black}generalisation~\cite{Bascone2018,Carollo2018,Avron2011,Bardyn2013,Huang2014,Viyuela2014,Viyuela2014a,Budich2015a,Linzner2016,Mera2017,Grusdt2017,Bardyn2018,Carollo2018a,Leonforte2019},
to account for the effect of temperature in systems at thermal
equilibrium or in out-of-equilibrium
scenarios~\cite{Magazzu2015,Magazzu2016,Spagnolo2018,Valenti2018,Spagnolo2018a,Guarcello2015,Guarcello2016,Spagnolo2016,Spagnolo2015}.
}

Moreover, much effort has been done to study the fault-tolerant
quantum computation via
topology\cite{Nayak2008,Preskill1997,Kitaev2003,Kitaev2009}. In this
context a main role is played by the Kitaev honeycomb model
\cite{Kitaev2006}, which possesses a rich phase structure allowing
for the presence of both Abelian and non-Abelian anyonic
excitations. Non-Abelian anyons are in fact a crucial building block
of topological quantum computing since one can perform unitary
operations by braiding these excitations. The model was also
analysed at finite temperature in Ref.~\cite{Bascone2018} by using
the mean Uhlmann curvature as a main tool. The analysis of finite
temperature phase transitions indeed is particularly important in
the quantum computing framework since it could help to understand
how the topological concepts can be used at finite temperature and
therefore better exploited in applications. The honeycomb model
under consideration shows a phase diagram containing gapped and
gapless phases. In particular, we introduce a
{\color{black}time-reversal symmetry breaking term}, such that the
system belongs to the symmetry-protected class $D${\color{black}.
The latter is } characterised by a {\color{black}$C=+1$} charge
conjugation symmetry and by the absence of time-reversal and chiral
symmetries \cite{Chiu2016}. Such an external perturbation allows for
the existence of non-Abelian excitations and opens a gap in an
otherwise gapless phase. The main motivation of this work however is
to introduce such a perturbation to analyse the Berry curvature of
the system. Accordingly, one of the main results of this paper is the
analysis of the Berry curvature of the Kitaev honeycomb model, both
numerically and analytically, carried out in the presence of a
time-reversal-symmetry-breaking perturbation. In particular, we will 
focus on the non-analytical behaviour of the curvature in the limit in 
which the above perturbation tends to zero. This procedure becomes 
necessary since the curvature is trivial in the vanishing perturbation 
case, which does not allow to gain any information on phase transitions 
eventually present in the system. On the contrary, extending the parameter 
manifold by including such a perturbation and then letting the coupling go to
zero allow to recover information about the topological phase
transitions.

The paper is organised as follows. In section~\ref{honey} we discuss
the spin honeycomb model and its phase diagram, by employing the
Fermionisation procedure introduced in Ref.~\cite{Kells2009}, which
has the advantage to provide a closed form of the ground state in a
BCS form. This technique permits to consider the system as a
two-band p-wave topological superconductor, allowing for more
convenient calculations. In section~\ref{berry} we carry out the
calculation of the Berry curvature for the ground state, which is
unique in the planar geometry. In particular, we calculate and
analyse the Berry curvature and its derivative in the vanishing
external perturbation limit, both analytically and numerically. The
analytic estimation is performed by expanding the curvature around
the Dirac points. Section~\ref{conclusions} contains the concluding
remarks.

\section{Honeycomb model}
\label{honey}

\begin{figure}[h!]
\centering
\includegraphics[width=0.6\linewidth]{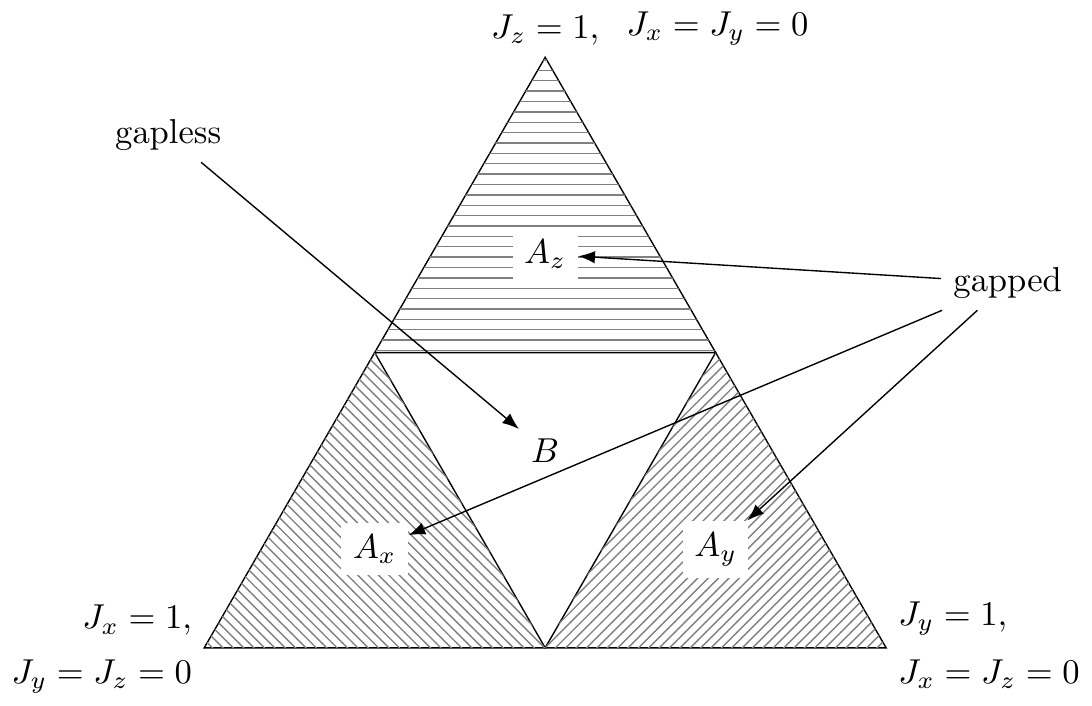}
\caption{Phase diagram of the honeycomb model: the triangle is the section of the positive octant by the plane $J_x+J_y+J_z=1$.}
\label{fig:phasediagram}
\end{figure}

In this paper we consider the Kitaev honeycomb model
\cite{Kitaev2006}, which consists of spin-$1/2$ particles arranged
on the vertices of a honeycomb lattice. This model can support a
rich variety of topological behaviors, depending on the values of
its couplings.\\
The Hamiltonian of the system can be written as follows
\begin{equation}
\label{ham}
H=-\sum_{\alpha \in \{x,y,z \}} \sum_{i,j}J_{\alpha}K^{\alpha}_{ij},
\end{equation}
\begin{figure}
\centering
\begin{minipage}{.4\textwidth}
  \centering
  \includegraphics[scale=0.9]{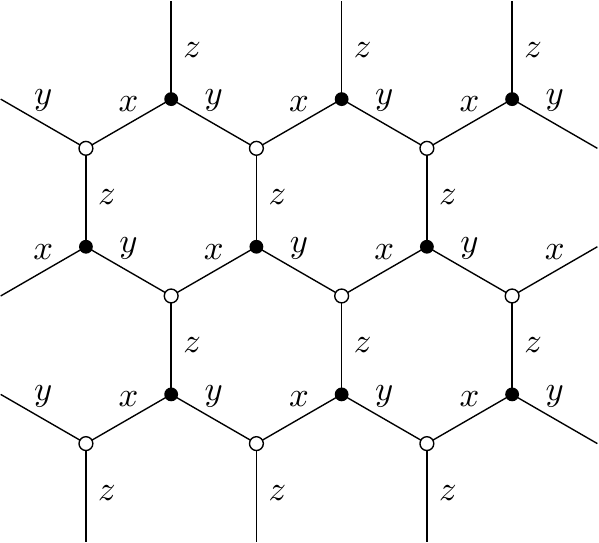}
  \captionof{figure}{Honeycomb lattice and link-types.}
  \label{fig:honeycomb}
\end{minipage}%
\begin{minipage}{.4\textwidth}
  \centering
  \includegraphics[scale=0.9]{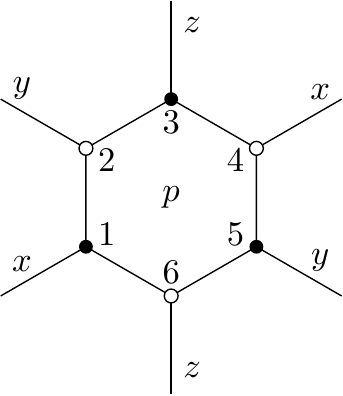}
  \captionof{figure}{Plaquette structure.}
  \label{fig:plaquette}
\end{minipage}
\end{figure}
with $K^{\alpha}_{ij}=\sigma_i^{\alpha} \sigma_j^{\alpha}$ denoting
directional spin interaction between $i$, $j$ sites connected by
$\alpha$-link (see Fig. \ref{fig:honeycomb}). $J_{\alpha}$ are the
dimensionless coupling coefficients of the two-body interaction and
the $\sigma_i^{\alpha}$ are the Pauli operators.\\
Products of the $K$ operators can be used to construct loops on the
lattice and any loop constructed in this way commutes with all other
loops and with the Hamiltonian. In particular, the shortest loop
symmetries are the plaquette operators
\begin{equation}
W_p=K_{12}K_{23}K_{34}K_{45}K_{56}K_{61}=\sigma_1^x \sigma_2^y \sigma_3^z \sigma_4^x \sigma_5^y \sigma_6^z,
\end{equation}
where $p \in \left\{1,2, \dotsc, m\right\}$ is a plaquette index, and $m$ is the number of plaquettes (Fig. \ref{fig:plaquette}). \\
These $W_p$ operators, which represent loops around single hexagons,
commute with each other and with the Hamiltonian.
Therefore they are integrals of motion and the Hilbert space of the
system can be divided into \emph{sectors}, each of
which is eigenspace of a different $W_p$. Each loop operator has
eigenvalues $\pm 1$, and plaquettes with $w_p=-1$ are said to carry
a vortex. Therefore, each sector corresponds to a particular choice
of the string of eigenvalues over all the plaquettes
$\left\{w_p\right\}|_{p \in \{1,2, \dotsc, m \}}$.\\
In this way, the Hamiltonian can be decomposed as a direct sum over
all the configurations:
\begin{equation}
\mathcal{H}=\bigoplus_{\left\{w_p \right\}}\mathcal{H}_{\left\{w_p\right\}}.
\end{equation}
Thus, one needs to find the eigenvalues of the Hamiltonian
restricted to a particular sector, and there are several ways to
exactly solve this problem. We use a Fermionisation approach first
developed in Refs.~\cite{Kells2009,Kells2008}. This technique
consists of a Jordan-Wigner (JW) Fermionisation, mapping
``hard-core'' bosons operators to Fermionic operators through string
operators. This procedure allows for an explicit construction of the
eigenstates of
the system, leading to a closed form of the groundstate.\\
A theorem by Lieb~\cite{Lieb1994} shows that the ground state of the
system must lie in the vortex-free sector, while
its degeneracy and form depend on the manifold
considered. By focusing on the vortex-free sector,
in a planar lattice geometry, we can take advantage of the
translational symmetry and use the Fourier transform to derive the
energy spectrum. By performing the JW transformation, we get the
following Bogoulibov-deGennes (BdG)-like Hamiltonian:
\begin{equation}
\label{bdghamiltonian}
H=\frac{1}{2}\sum_{\textbf{q}}\left(C^{\dagger}_{\textbf{q}}, \, C_{-\textbf{q}} \right)
 H_{\textbf{q}} \begin{pmatrix}
C_{\textbf{q}} \\
C^{\dagger}_{-\textbf{q}}
\end{pmatrix},
\end{equation}
where,
\begin{equation}
\label{22ham}
H_{\textbf{q}} \equiv \begin{pmatrix}
\xi_{\textbf{q}} & \Delta_{\textbf{q}} \\
\Delta^*_{\textbf{q}} & -\xi_{\textbf{q}}
\end{pmatrix},
\end{equation}
with
\begin{equation}
\begin{aligned}
{} & \xi_{\textbf{q}}=2J_{x}\cos q_x+2J_{y}\cos q_y+2J_z, \\ &
\Delta_{\textbf{q}}=i\beta_{\textbf{q}}=2 i J_{x} \sin q_x+2 i J_{y} \sin q_y.
\end{aligned}
\end{equation}
Here we use a Cartesian basis where $\textbf{q} \equiv \left(q_x,
q_y \right)$.\\
Thus, the Kitaev honeycomb model is mapped into a spinless Fermionic
BdG Hamiltonian. The Hamiltonians $H_{\textbf{q}}$ can be
diagonalised via Bogoliubov rotation of the mode operators, and the
diagonalised Hamiltonian becomes
\begin{equation}
\label{diagoham}
H=\sum_{\textbf{q}}\epsilon_{\textbf{q}}\left(b^{\dagger}_{\textbf{q}}b_{\textbf{q}}-\frac{1}{2} \right),
\end{equation}
whose ground state has the BCS form
\begin{equation}
\ket{\Psi_0}=\prod_{\textbf{q}}\left(u_{\textbf{q}}+v_{\textbf{q}}C^{\dagger}_{\textbf{q}} C^{\dagger}_{-\textbf{q}}\right) \ket{0}.
\end{equation}
From the dispersion relation
\begin{equation}
\epsilon_{\textbf{q}}=\sqrt{\xi^2_{\textbf{q}}+\left|\Delta_{\textbf{q}} \right|^2}=\sqrt{\xi^2_{\textbf{q}}+\beta^2_{\textbf{q}}}
\end{equation}
it is possible to find out the phase diagram structure of the system.\\
Indeed it can be easily checked that the following triangular
inequalities
\begin{equation}\label{Bphase}
\left|J_{x} \right| \leq \left|J_{y} \right| + \left|J_{z} \right|, \quad \left|J_y \right| \leq \left|J_x \right| + \left|J_z \right|,  \quad \left|J_z \right| \leq \left|J_x \right| + \left|J_y \right|,
\end{equation}
if satisfied, identify the gapless regions of the spectrum. In
Fig.~\ref{fig:phasediagram} we explicitly show the above triangular
condition in the positive octant ($J_x, J_y, J_z \geq 0$). The
graphical representation in the other octants can
be easily derived by symmetry. The triangular region in the phase
diagram determined by such triangular conditions is named gapless B
phase, while the other three regions (equivalent to
each other) are indicated as gapped A phases.\\
\section{Berry curvature analysis in the $\kappa \rightarrow 0$ limit}
\label{berry}

In this section, in view of extending the results
obtained in Ref.~\cite{Bascone2018}, we calculate the Berry
curvature, closely
following the procedure used in that work.
In particular, we focus on the vertex-free configuration in a planar
geometry, to take into account only a
single ground state and therefore to get an Abelian
Berry curvature. Note that in general for the current analysis it is
not necessary to embed the system on a torus.\\
The $2 \times 2$ Hamiltonian in Eq.~(\ref{22ham}) can be rewritten
explicitly as
\begin{equation}
\label{hampaul}
H_{\textbf{q}}=\textbf{h}(J) \cdot \sigma,
\end{equation}
where $\textbf{h}(J) \equiv \left(0,\, -\beta_{\textbf{q}}, \,
\xi_{\textbf{q}} \right)$, and $\sigma$ are the Pauli matrices. In
this form, the spectral Berry curvature can be
easily computed by means of the relation
\begin{equation}
\label{curvham}
\mathcal{F}_{ij}=\frac{1}{2h^3}\left[\left(\partial_i \textbf{h}\right) \times \left(\partial_j \textbf{h}\right) \right] \cdot \textbf{h},
\end{equation}
where $h:=\left|\textbf{h} \right|=\epsilon_{\textbf{q}}$ and $\partial_{i}:= \partial/\partial J_{i}$.\\
It is straightforward to check that this curvature appears to be
zero everywhere. This can be deduced as a result of time-reversal
(TR) and parity (P) symmetries of the model. Namely, one can note
that the $\textbf{h}(J)$ vector and all its derivatives are entirely
contained in the $(y,z)$ plane.

As discussed in introduction, adding a TR and/or P symmetry-breaking
term in the Hamiltonian, for instance by means of an external
magnetic field, results in a gap opening in the $B$ phase. This
condition allows for the existence of non-Abelian anyonic
excitations. Explicitly, one can add a three-body interaction term
of the form \cite{Kells2014}
\begin{equation}
\label{hint}
H_{\text{int}}=-\kappa \sum_{\textbf{q}} \left(\sigma_1^x \sigma^y_6 \sigma^z_5+\sigma_2^z \sigma_3^y \sigma_4^x+\sigma_1^y\sigma_2^x \sigma_3^z+\sigma_4^y \sigma_5^x \sigma_6^z \right),
\end{equation}
where $\kappa$ is the three-body external coupling, playing the role
of an "effective magnetic field".

\begin{figure}[h]
\centering
\includegraphics[width=0.6\linewidth]{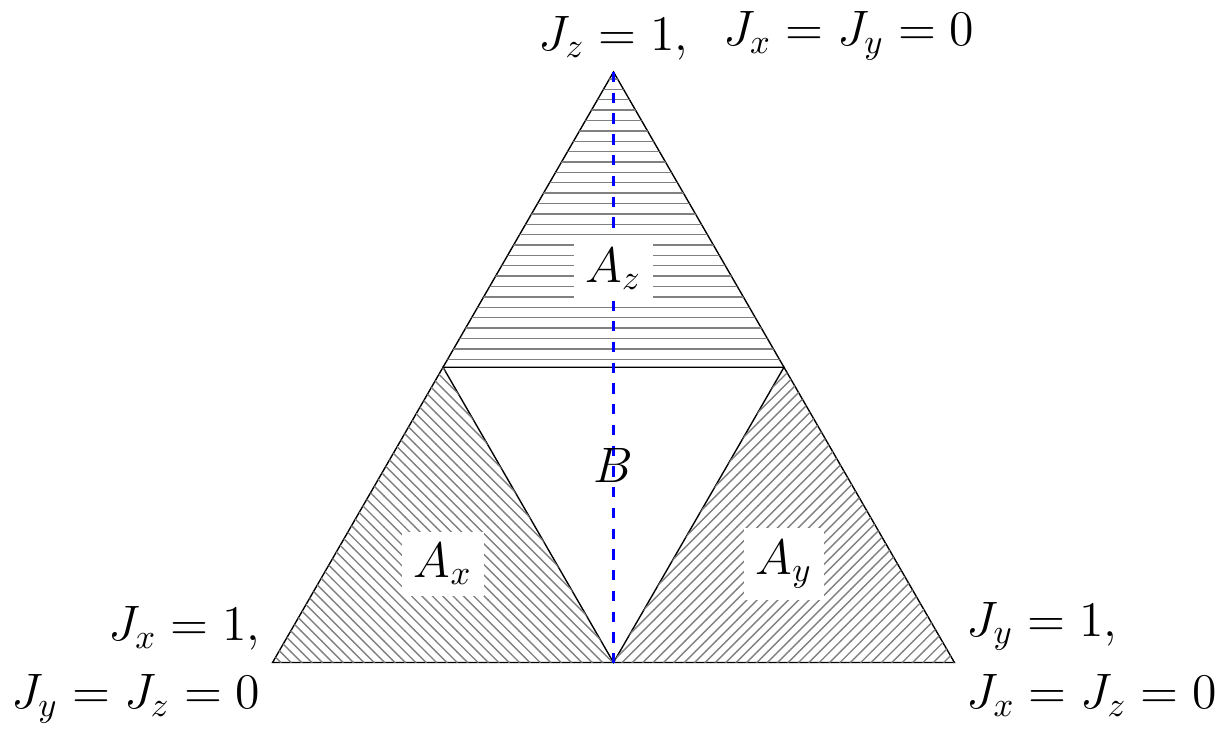}
\caption{Phase diagram: the blue dashed line, taken as the line on which the Berry curvature is explored, is parametrised by $J_x=J_y=J$.}
\label{fig:phasediagramlines}
\end{figure}

\begin{figure}[h!]
\centering
\includegraphics[width=0.45\linewidth]{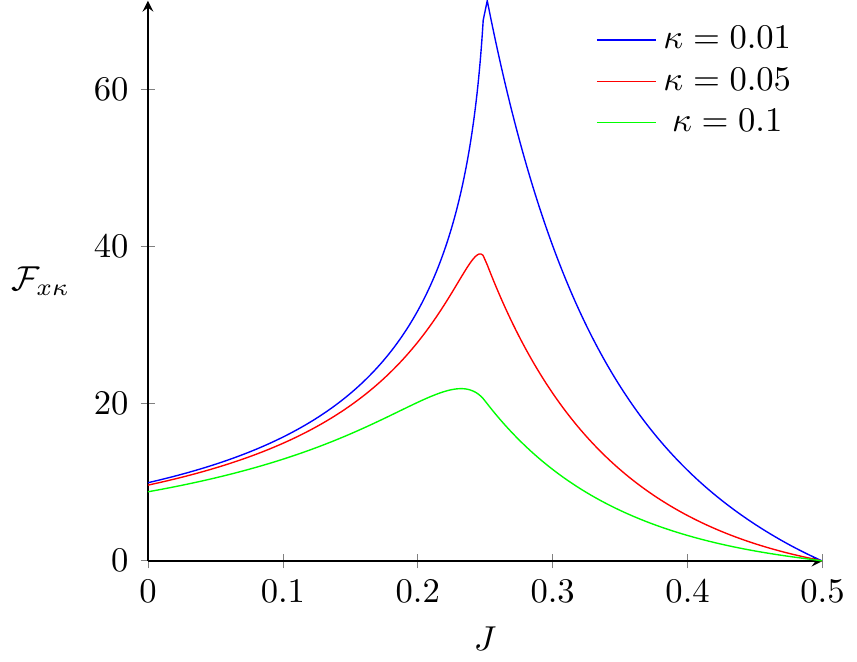}
\caption{$\mathcal{F}_{x \kappa}$ component of the Berry curvature as a function of $J$ along the evolution line $J_x=J_y=J$, $J_z=1-2J$, with external coupling values $\kappa=0.01, \, 0.05, \, 0.1$.}
\label{fig:berry}
\end{figure}

The Hamiltonian $H_{\textbf{q}}$ in Eq. (\ref{22ham}) remains of the
same form, provided that a real part is added to
$\Delta_{\textbf{q}}$, that is
$\Delta_{\textbf{q}}=\alpha_{\textbf{q}}+i\beta_{\textbf{q}}$, with
\begin{equation}
\alpha_{\textbf{q}}=4\kappa\left[\sin q_x-\sin q_y \right].
\end{equation}
The diagonalised form of this Hamiltonian is then exactly the same as in Eq.(\ref{diagoham}), but with
\begin{equation}
\epsilon_{\textbf{q}}=\sqrt{\xi^2_{\textbf{q}}+\left|\Delta_{\textbf{q}} \right|^2}=\sqrt{\xi^2_{\textbf{q}}+\alpha^2_{\textbf{q}}+\beta^2_{\textbf{q}}}.
\end{equation}
We can still write $H_{\textbf{q}}$ in the form of Eq.(\ref{hampaul}), but with a slightly different vector $\textbf{h}(J)\equiv \left(\alpha_{\textbf{q}},\, -\beta_{\textbf{q}}, \, \xi_{\textbf{q}} \right)$, and calculate again the spectral curvature.
Of course, one should  extend the $3$-dimensional parameter manifold to a $4$-dimensional one to include the extra parameter $\kappa$.\\
We find that the only non-vanishing components of the curvature
in~Eq.(\ref{curvham}) are the $\mathcal{F}_{i
\kappa}=-\mathcal{F}_{\kappa i}$, $i \in \{x,y,z \}$, which are
explicitly given by
\begin{align*}
{} & \mathcal{F}_{x\kappa, \textbf{q}}=\frac{\left[\sin q_x -\sin q_y \right]}{2\epsilon^3_{\textbf{q}}}\left[\xi_{\textbf{q}}\sin q_x -\beta_{\textbf{q}}\cos q_x \right], \\ &
\mathcal{F}_{y\kappa, \textbf{q}}=\frac{\left[\sin q_x -\sin q_y \right]}{2\epsilon^3_{\textbf{q}}}\left[\xi_{\textbf{q}}\sin q_y -\beta_{\textbf{q}}\cos q_y \right], \\ &
\mathcal{F}_{z\kappa, \textbf{q}}=-\frac{\left[\sin q_x -\sin q_y \right]}{2\epsilon^3_{\textbf{q}}}\beta_{\textbf{q}}.
\end{align*}
In order to obtain the total curvature, the spectral curvature $\mathcal{F}_{i \kappa}$ needs to be summed over all quasi-momenta $\textbf{q}$ (or, in the thermodynamic limit, integrating over $d\textbf{q}$).\\
Without loss of generality, we choose the octant with $J_i\geq0 \,
\, \, \, \forall i \in \{x,y,z\}$. The three gapped phases $A_i$
correspond to the region $\{J_i > J_j+J_k, \, i \neq j \neq k \in
\{x,y,z\}\}$ in the parameter space.  The $B$ phase is instead
realised by the conditions~(\ref{Bphase}). The four phases are
separated by quantum phase transition lines on which one of the
$J_i$ is equal to the sum of the other two (see Fig.
\ref{fig:phasediagram}). A TR-P breaking perturbation (for instance
the term in Eq. (\ref{hint}) with $\kappa \neq 0$) ) opens a gap in
the otherwise gapless phase $B$. This would make both the $A$s and
the $B$ phases gapped, but still with different properties: indeed,
$A$ phases host Abelian excitations, whereas the low energy
excitation of the $B$ phase satisfy non-Abelian anyonic statistics.
Notice that, in the chosen octant, the two phases are separated by
the plane $J_x+J_y+J_z=1$, and independently of the phase we are in,
the couplings have to satisfy such a normalisation condition. To
explore the behaviour of the Berry curvature in the different phases
and in particular on the transition lines between them, we can
choose to study, without loss of generality, the system along the
$J_x=J_y$ line, cutting vertically the triangle diagram (blue dashed
line in Fig. \ref{fig:phasediagramlines}).
This choice of cut line allows one to explore the
dependence of the curvature in the $A_z$ and $B$ phases on $J_z$,
with a special focus on the critical line at $J_z=\frac{1}{2}$.
Under these conditions we can use $J_x=J_y=J$ and, because of the
normalisation relation $J_z=1-2J$, the curvature components are
expressed as functions of a single parameter $0 \leq J \leq
\frac{1}{2}$ along this line (the transition at $J_z=\frac{1}{2}$ is
realised at $J=\frac{1}{4}$).\\
Along this evolution line, the terms appearing in the expressions
for the curvature components can be simplified as follows:
\begin{align}
\alpha_{\textbf{q}}=&4\kappa\left[\sin q_x-\sin q_y \right] \nonumber\\
\beta_{\textbf{q}}= &2J\left(\sin q_x +\sin q_y \right), \nonumber\\
\xi_{\textbf{q}}= &2J\left(\cos q_x +\cos q_y\right)+2-4J, \label{eq:alpha-beta}\\
\epsilon_{\textbf{q}}=&\sqrt{\xi^2_{\textbf{q}}+\alpha^2_{\textbf{q}}+\beta^2_{\textbf{q}}}=\nonumber\\
=&\left\{8J^2[\cos(q_x-q_y )+1]+16\kappa^2[\sin  q_x- \sin q_y ]^2+\right.\nonumber\\
&\left.+(2-4J)[2+4J(\cos q_x +\cos q_y -1)]\right\}^{1/2}\nonumber,
\end{align}
so that the Berry curvature components in the thermodynamic limit have the form
\begin{equation*}
\mathcal{F}_{i \kappa}(J)=\int_{-\pi}^{\pi}\int_{-\pi}^{\pi} dq_x dq_y \, \mathcal{F}_{i \kappa, \textbf{q}}(J),
\end{equation*}
with
\begin{align}
\mathcal{F}_{x \kappa, \textbf{q}}=& 8 \left(\sin q_x-\sin q_y \right) \times\nonumber\\&\times\left[J \sin(q_x-q_y)+\left(1-2J \right)\sin q_x \right]\epsilon_{\textbf{q}}^{-3}, \nonumber\\
\mathcal{F}_{y \kappa, \textbf{q}}=& 8\left(\sin q_x-\sin q_y \right) \times\label{comp}\\&\times\left[J \sin(q_y-q_x)+\left(1-2J \right)\sin q_y \right]\epsilon_{\textbf{q}}^{-3},\nonumber\\
\mathcal{F}_{z \kappa, \textbf{q}}=& 8J\left(\sin^2  q_y-\sin^2 q_x \right) \epsilon_{\textbf{q}}^{-3}.\nonumber
\end{align}
It is easy to see that only one of the above expressions is
independent since $\mathcal{F}_{x \kappa}(J)=-\mathcal{F}_{y
\kappa}(J)$ and $\mathcal{F}_{z \kappa}(J)=0$. This means that we
can limit our analysis just to the $\mathcal{F}_{x \kappa}(J)$
component. As it is discussed in \cite{Bascone2018}, this is a
consequence of the specific symmetry of the chosen cut-line.\\
The numerical result of the integration along the line with
$J_x=J_y=J$ for different values of $\kappa \neq 0$ is shown in Fig.
(\ref{fig:berry}). It is interesting to note that the function is
peaked close to the transition line, at $J=\frac{1}{4}$, while it is
regular enough over the whole region $0 \leq J \leq \frac{1}{2}$.
For $\kappa \neq 0$ we can expect that eventual
criticalities may not be evidenced by the Berry curvature, while
they are surely identified by the Chern number.
It is also worth noting that the vertical line in the phase diagram
(see Fig.~\ref{fig:phasediagramlines}) is travelled downward, so
that the $A_z$ phase is covered for $0\leq J < \frac{1}{4}$ while
the $B$ phase is covered for $\frac{1}{4} < J \leq \frac{1}{2}$.

\begin{figure}[h!]
\centering
\includegraphics[width=0.55\linewidth]{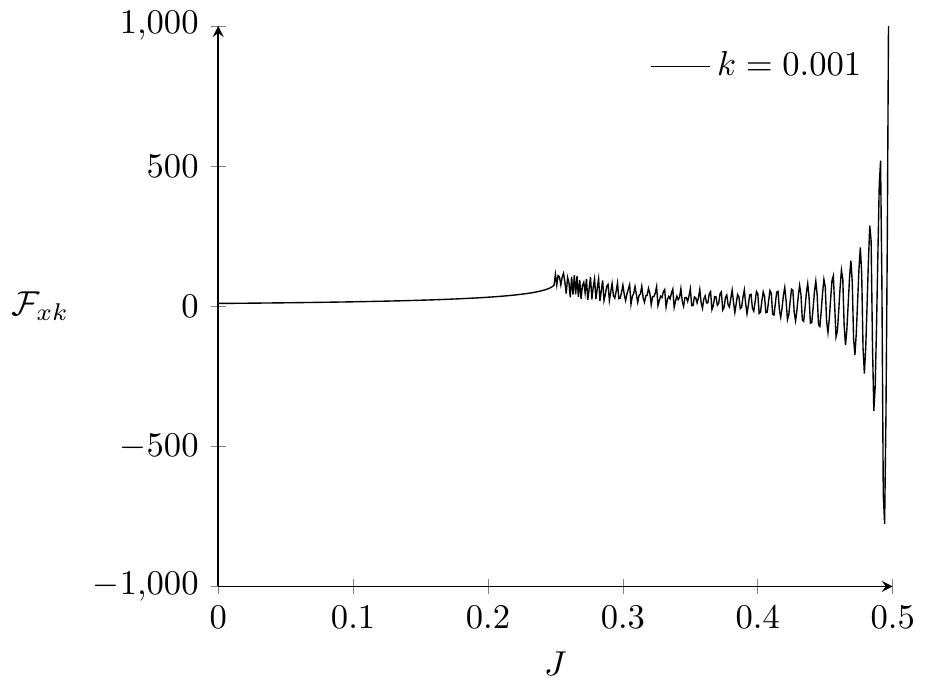}
\caption{$\mathcal{F}_{x \kappa}$ component of the Berry curvature
as a function of $J$ along the evolution line $J_x=J_y=J$,
$J_z=1-2J$, with $\kappa=0.001$ to resemble the $\kappa \rightarrow
0$ case.} \label{fig:berryk0}
\end{figure}

The Berry curvature peak gets higher as $\kappa$ decreases to zero.
This can be explained on account of the inverse dependence of the
Berry curvature on the gap, which, in turn, tightens as $\kappa$
decreases.
To analyse the $\kappa \rightarrow 0$ case, we study the Berry curvature analytically, by estimating the integrals around the Dirac points. This approach is justified by the fact that the dominant contribution to the Berry curvature comes from the regions close to the Dirac points.\\ This analytical approximation is validated by a numerical analysis performed for small enough values of $\kappa$, yet large enough to avoid numerical instabilities.\\
The analytical approach entails finding the minima of the energy
spectrum around which the integrand in Eq. (\ref{comp}) is expanded
in series (we consider again only the $\mathcal{F}_{xk}$ component).
The position of the minima crucially depends on the phases of the
model, i.e. whether the coupling $J$ is larger or smaller than the
critical value $J=1/4$. We distinguish the expansion in two separate
cases.
From the analysis of the function $\epsilon_{\textbf{q}}$ in the $J
\in (\frac{1}{4}, \frac{1}{2}]$ region, it follows that two minima
are found for the following values of the momentum components
\begin{equation}
q^*_x=-q_y^*=\pm \arccos\left(\frac{1-\frac{1}{2J}}{1-\left(\frac{2\kappa}{J} \right)^2} \right).
\end{equation}
By performing a second order expansion of the integrand function
$\mathcal{F}_{x \kappa, \textbf{q}}$ around these minima and using
the eigenvalues of the Hessian matrix
\begin{equation}
e_1=16\left(J-\frac{1}{2} \right)^2, \quad e_2=16\left(J-\frac{1}{4}
\right),
\end{equation}
along the minimum eigendirections we are left to compute the following integral:
\begin{equation}
\int_{-R}^R  \int_{-R}^R dx \, dy \frac{N_0+N_1x^2+N_2y^2}{\left(A^2+B^2 x^2+C^2 y^2\right)^{3/2}}=I_0+I_1+I_2,
\end{equation}
with
\begin{equation}
\begin{aligned}
{} & N_0=-\frac{8}{J^2}\left(J-\frac{1}{4} \right)\left(1-2J \right)\left(\frac{2\kappa}{J} \right)^2, \\ &
N_1=\frac{-40}{J^2}\left(\frac{1}{2}-J \right)\left(J-\frac{1}{4} \right), \quad N_2=\frac{8}{J^2}\left(\frac{1}{2}-J \right)\left(J-\frac{1}{4} \right),
\\ &
A=\frac{8\kappa}{J}\sqrt{J-\frac{1}{4}}, \quad B=4\left(\frac{1}{2}-J \right), \quad C=4\sqrt{J-\frac{1}{4} }.
\end{aligned}
\end{equation}
We also used the fact that the cross terms in the expansion are odd and they do not contribute in the symmetric integration region. The integration variables $x$ and $y$ are the eigencoordinates, i.e. the momentum variables in the basis where the Hessian is diagonal. The finite integration radius $R$ is taken to enclose the minima and its explicit value is not important for the estimate.
It is not hard to see that the contribution coming from $I_0=\int_{-R}^R  \int_{-R}^R dx \, dy \frac{N_0}{\left(A^2+B^2 x^2+C^2 y^2\right)^{3/2}}$ vanishes in the $\frac{\kappa}{J} \rightarrow 0$ limit, while for the other two contributions we find, in the same limit,
\begin{align*}
&\mathcal{F}_x=\lim_{\frac{\kappa}{J} \to 0} \left(I_1+I_2 \right) \\ &\propto\frac{1}{J^2}\left[\frac{\log\left(z+\sqrt{1+z^2} \right)}{z}-5z^2\log\left(\frac{1}{z}+\sqrt{1+\frac{1}{z^2}} \right) \right],
\end{align*}
with $z=\frac{\sqrt{J-\frac{1}{4}}}{\frac{1}{2}-J}$.
We note that in the $J \rightarrow \frac{1}{4}$ limit the Berry curvature is finite, which is in agreement with the numerical analysis.\\
However, even if there is no criticality, the Berry curvature still
gives information about the different phases of the system: it can
be seen numerically that for very small values of $\kappa$,
resembling the $\kappa \rightarrow 0$ limit, very different
behaviors are found below and above the transition line
$J=\frac{1}{4}$. Namely, rapid oscillations appear in the
non-trivial phase similarly to the behavior found in
Ref.~\cite{Yang2008} for the fidelity susceptibility, as shown in
Fig.~\ref{fig:berryk0}, explicitly revealing the two different
topological phases.
Such oscillations however seems to be related to the finite system
size used for numerical analysis. One can indeed observe that the
height of the peaks increases with the system size.
\begin{figure}[h!]
\centering
\includegraphics[width=0.55\linewidth]{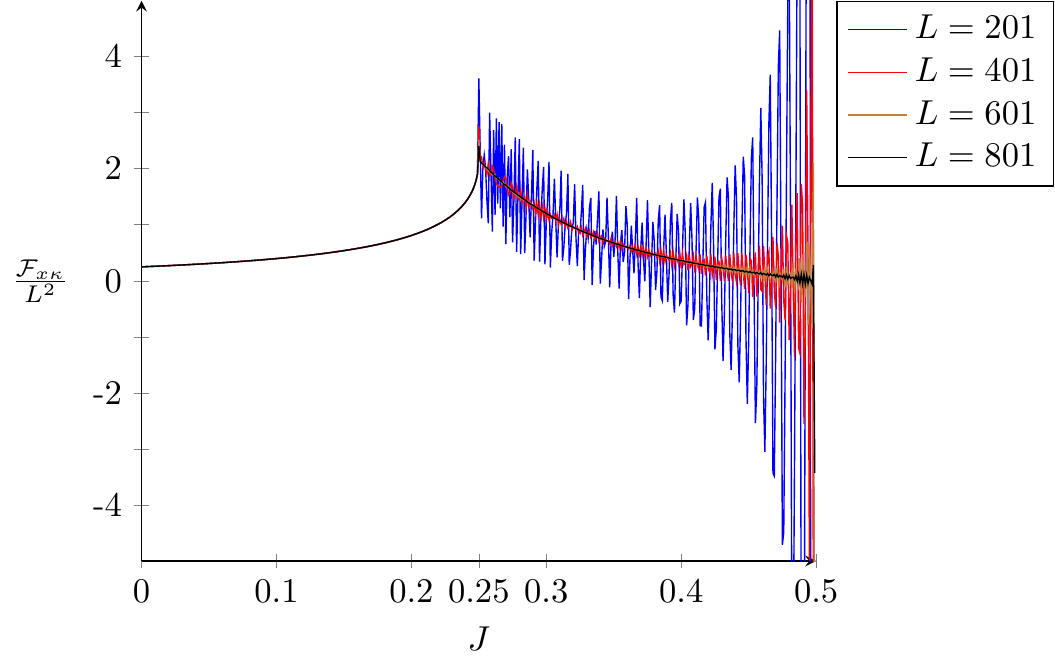}
\caption{$\mathcal{F}_{x \kappa}$ component of the Berry curvature as a function of $J$ along the evolution line $J_x=J_y=J$, $J_z=1-2J$, with $\kappa=0.001$ to resemble the $\kappa \rightarrow 0$ case.}
\label{fig:curvosc}
\end{figure}
Since the Berry curvature does not show any criticalities, it is
relevant to analyse also its first derivative (w.r.t. the parameter
$J$). This analysis allows to estimate the derivative of the
curvature, providing the following result:
\begin{equation}
\partial_J \mathcal{F}_{x\kappa} \propto \frac{\log\left(J-\frac{1}{4} \right)}{J^2},
\end{equation}
which diverges to $-\infty$ in the $J \rightarrow \frac{1}{4}^+$
limit, indicating the presence of a criticality (it is worth
reminding that at this stage we are only considering the region at
the right of the transition point for $J$: $(\frac{1}{4},
\frac{1}{2}]$). Now we extend this analysis to the left of the
transition point. In the $J \in [0,\frac{1}{4})$ region the only
minimum in the momentum space is the point $(\pi, \pi)$, around
which the curvature is expanded. Following the same
procedure as before, we calculate the eigenvalues of the Hessian
matrix in $(\pi, \pi)$:
\begin{equation}
\begin{aligned}
{} & e_1= -8\left(4J^{2}-J-8 k^{2}\right), \quad e_2= -8\left(2 J^{2}-J\right),
\end{aligned}
\end{equation}
where we use the condition $J > 2\kappa$, which is compatible since
we are interested in the $\frac{\kappa}{J} \rightarrow 0$ limit.
Expanding $N(\textbf{q}; J)$ up to the second order around $(\pi,
\pi)$ we have
\begin{equation}
N(\textbf{q}; J)=-8(q_x-q_y)\left[(3 J-1) q_x-J q_y\right],
\end{equation}
while the second order expansion of the energy is given by
\begin{equation}
\epsilon^2(\textbf{q};J, \kappa)=4\left[J^{2}\left(-3 q_x^2+2 q_x q_y-3 q_y^2+16\right)+J\left(q_x^2+q_y^2-8\right)+4 k^2(q_x-q_y)^2+1\right].
\end{equation}
Under a suitable rotation the Hessian becomes a
diagonal matrix. By rewriting the above function with respect to the rotated variables, we obtain
\begin{equation}
N(\textbf{q}; J)=\left(8-32J \right) q_x^2 + \eta \, q_x q_y
\label{N_diag}
\end{equation}
and
\begin{equation}
\epsilon^2(\textbf{q};J, \kappa)=4\left[\left(-4J^2+8\kappa^2+J \right)q_x^2+\left(J-2J^2 \right) q_y^2+16\left(J-\frac{1}{4} \right)^2\right]
\end{equation}
In Eq.~(\ref{N_diag}) we do not need to specify the
value of $\eta$ since, under the symmetric integration domain, the
mixed $q_x q_y$ term coming from the numerator $N$ vanishes.
Hence, we are left to compute the following integral:
\begin{equation}
\int_{-R}^R \int_{-R}^R dx \,  dy \frac{-N_0 \, q_x^2}{\left(A+Bx^2+Cy^2\right)^{3/2}}
\end{equation}
with
\begin{equation}
\begin{aligned}
{} & N_0=32\left(\frac{1}{4}-J \right), \quad A=16\left(\frac{1}{4}-J \right)^2, \\ &
B=4\left(-4J^2+8\kappa^2 +J\right), \quad C=2J\left(\frac{1}{2}-J \right).
\end{aligned}
\end{equation}
The integral computation is straightforward, and leads to the
following result for the Berry curvature estimation in the
$[0,\frac{1}{4})$ region in the $\kappa \rightarrow 0$ limit:
\begin{equation}
\begin{aligned}
\mathcal{F}_{|\kappa \rightarrow 0} {} & \propto -\frac{1}{J^{3/2}\sqrt{\frac{1}{4}-J}}\log \left(\frac{4R \sqrt{J\left(\frac{1}{4}-J \right)}+\sqrt{16(\frac{1}{4}-J)^2+J(5-18J)R^2}}{\sqrt{16(\frac{1}{4}-J)^2+2R^2J(\frac{1}{2}-J)}} \right) \\ & +\frac{2}{J^2} \sqrt{\frac{(\frac{1}{2}-2J)}{(\frac{1}{2}-J)}} \arctan\left(\frac{\sqrt{32(\frac{1}{4}-J)(\frac{1}{2}-J)}JR^2}{\sqrt{16(\frac{1}{4}-J)^2+R^2J(5-18J)}} \right).
\end{aligned}
\end{equation}
\begin{figure}[h!]
\centering
\includegraphics[width=0.48\linewidth]{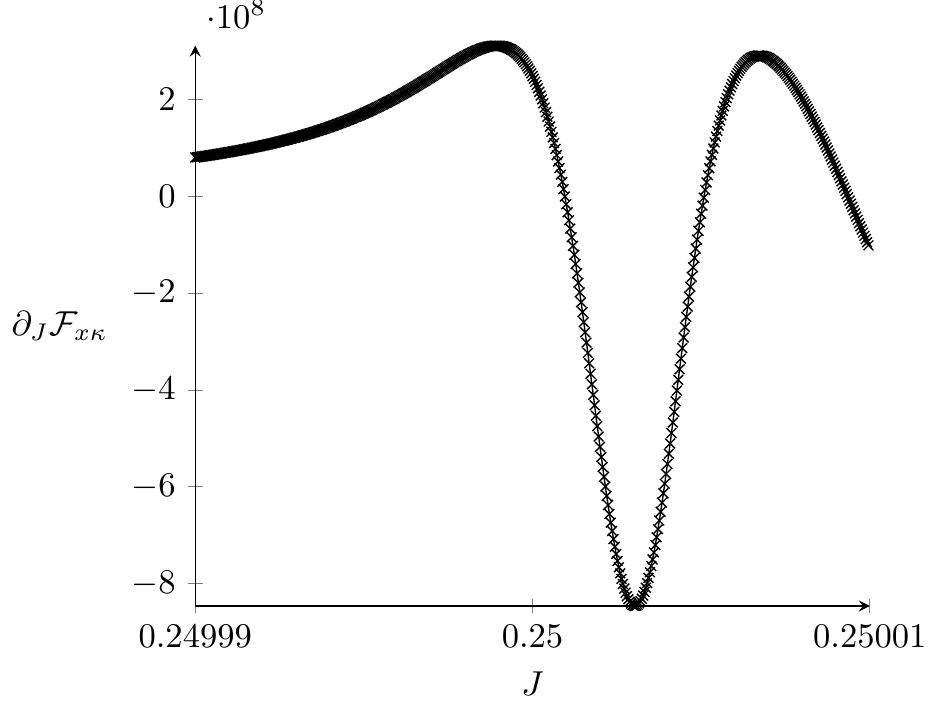}
\caption{First derivative of the $\mathcal{F}_{x \kappa}$ component of the Berry curvature with $\kappa=0.001$ in the vicinity of the transition line.}
\label{fig:derivative}
\end{figure}
It is easy to see that the latter expression is finite in the $J
\rightarrow {\frac{1}{4}}^-$ limit. However, just as in the right
limit case, we also analyse the first derivative behaviour. The
expression for the derivative is the following:
\begin{equation}
\partial_{J} \mathcal{F}_{|\kappa \rightarrow 0} \propto \frac{ \log \left(R\varphi_1(J) \right)}{\sqrt{\varphi_2 (J)}}-4R\frac{\psi_1(J)}{\left(J-2 J^2\right)^2 \sqrt{\psi_2(J)}},
\end{equation}
with
\begin{equation}
\begin{aligned}
\varphi_1(J) {} & =  R^4 J^2 (5-36 J) (1-4 J) (1-2 J) \\ & -(1-4
J)^2 \big\{J \left[3 R^2 (8 J-1) (18 J-5)-8 \big(32 J^2-22
J+5\big)\right] +3\big\} \\ & +R^2 J \left\{2 J \left[R^2 (18 J-5) (
24J^2-15J+2)+(8 J-3) (112J^2-48J+7)\right]+1\right\} \\ & -4 R (1-8
J)  \sqrt{\left\{J^2 \left[R^2 (5-18 J)+16 J-8\right]+J \right\}
(1-4 J)^5} \\ & -2 R^3 (3-8 J) \sqrt{\left\{J^4 \left[R^2 (5-18
J)+16 J-8\right]+J^3\right\}(1-4 J)^3},
\end{aligned}
\end{equation}
\begin{equation}
\begin{aligned}
\varphi_2(J) {} & = (1-4 J)^3 J^5 \left[1-\left(R^2-8\right) (2 J^2-J)\right]^3 \left[J  \left(R^2 (5-18 J)+16 J-8\right)+1 \right],
\end{aligned}
\end{equation}
\begin{equation}
\begin{aligned}
\psi_1(J) {} & = \frac{R^2 J \sqrt{8 J^2-6 J+1} \left\{J \left[R^2 (1-6 J) (48 J^2-30J+5)+32 J (8 J^2-9J+4) -26\right]+2\right\}}{4 \sqrt{J \big(R^2 (5-18 J)+16 J-8\big)+1} \left\{J \left[4 R^4 J\big(8 J^2-6 J+1\big)+R^2 (5-18 J)+16 J-8\right]+1\right\}} \\ & -\frac{1}{2} (8J^2-5J+1) \arctan\left(\frac{2 R^2 J \sqrt{8 J^2-6 J+1}}{\sqrt{J \left[R^2 (5-18 J)+16J-8\right]+1}}\right),
\end{aligned}
\end{equation}
\begin{equation}
\begin{aligned}
\psi_2(J) {} & =  \frac{(4 J-1)}{(2 J-1)} \arctan\left(\frac{2 R^2 J \sqrt{8 J^2-6 J+1}}{\sqrt{J \left[R^2 (5-18 J)+16 J-8\right]+1}}\right).
\end{aligned}
\end{equation}
It can be seen that this derivative diverges to $-\infty$ in the $J
\rightarrow {\frac{1}{4}}^-$ limit, in agreement with the numerical
result (see Fig. (\ref{fig:derivative})). We observe that the
position of the criticality is not exactly at $J=\frac{1}{4}$ but
slightly shifted on the right. This is due to the finite system size
used in the numerical analysis. As expected, one can show that in the
limit of size tending to infinity
the criticality tends towards $J=\frac{1}{4}$.\\
Therefore, the analysis of the Berry curvature at $\kappa
\rightarrow 0$ shows a critical behaviour, revealing a topological
phase transition at $J \rightarrow \frac{1}{4}$. The
"detection" of this criticality would not be possible without the
expansion procedure we carried out in the parameter space.

\section{Conclusions}
\label{conclusions}

After briefly reviewing the Kitaev honeycomb model, we used a
{\color{black}suitable} Fermionisation technique to map our
Hamiltonian to a BdG one, obtaining explicit relations for the
relevant quantities we were interested in. In particular, we assumed
a translationally symmetric condition, by considering the
vortex-free sector of the model on an infinite plane. In Sec.
\ref{berry} we have calculated the Berry curvature by assuming an
expanded parameter manifold, which included an external effective
magnetic field. This perturbation changes the class of the model
from an intrinsic topological material to a symmetry protected one.
This allowed to get an analytical headway for the calculation of the
Berry curvature in the $\kappa \rightarrow 0$ limit, in which the
curvature would be otherwise identically zero. We estimated the
Berry curvature in the $\kappa \rightarrow 0$ limit by expanding
around the relevant Dirac points. We found no criticalities,
although this procedure provides information about
the phase transitions of the system due to the appearance of rapid
oscillations in the non-trivial phase. To better investigate the
origin of these oscillations, we calculated the first derivative of
the Berry curvature, which showed a divergence that clearly signals
a phase transition. Therefore, the analysis of the Berry curvature
in the $\kappa \rightarrow 0$ limit shows a criticality in the
transition line that was not possible to estimate without an
appropriate expansion in the parameter space.


\section*{Acknowledgments}
This work was supported by the Grant of the Government of the Russian Federation, contract No. 074-02-2018-330 (2). 
We acknowledge also partial support by Ministry of Education, University and Research of the Italian government.

\vspace{180pt}

\providecommand{\newblock}{}
\bibliographystyle{iopart-num}


\bibliography{ref}
\end{document}